\documentclass[a4paper,11pt]{article}
\usepackage{pos}

\title{Interaction between two hadrons in lattice QCD}
\author*[a]{Sinya Aoki}
\affiliation[a]{Center for Gravitational Physics and Quantum Information, Yukawa Institute for Theoretical Physics, Kyoto University,
  Kitashirakawa Oiwakecho, Sakyo-ku, Kyoto 606-8502, Japan}

\emailAdd{saoki@yukawa.kyoto-u.ac.jp}
\abstract{I report recent developments on hadron interactions in lattice QCD by the HAL QCD method.
As an introduction, I summarize the lattice community's consensus on the absence of the deeply bound dinucleons  at heavier pion masses.
We then present 4 results by the HAL QCD method using $2+1$ flavor QCD gauge configurations at $m_\pi \simeq 146$ MeV and $a\simeq 0.0846$ fm, which are the $\Lambda\Lambda-N\Xi$ interactions,  $N\Omega$ dibaryon, the tetra-quark state $T_{cc}$ and the $N-\phi$ interaction with the 2-pion tail.  A brief summary follows.}

\FullConference{The 11th International Workshop on Chiral Dynamics (CD2024)\\
 26-30 August 2024\\
Ruhr University Bochum, Germany\\}

\begin{document}
\begin{flushright}
YITP-25-33
\end{flushright}

\maketitle

\section{Introduction}
In lattice QCD, there are two methods utilized to extract interactions between hadrons,
one is the finite volume (FV)  method, which extracts  scattering phase shifts from spectra of two hadrons in finite boxes through L\"uscher's finite volume formula\cite{Luscher:1990ux}, the other is the HAL QCD method,
which first extracts a potential (interaction kernel) from the NBS (Nambu-Bethe-Salpeter) wave function and then
obtains scattering phase shifts by solving Schr\"odinger equation with the potential\cite{Ishii:2006ec,Aoki:2009ji,Aoki:2012tk}.

It is always better to have two different methods  for crosschecks.
Indeed, in the past, there appeared so called ``the NN controversy'' in the community:
While the FV method claimed that both deuteron and dineutron are bound at heavier pion masses\cite{Yamazaki:2012hi,NPLQCD:2012mex,Yamazaki:2015asa,Orginos:2015aya,Berkowitz:2015eaa}, 
 the HAL QCD potential method found that both are unbound there\cite{Ishii:2012ssm,Inoue:2011ai}.
 
 Since then, there appeared more results using the FV method  from the community to resolve this controversy. 
At lattice 2023, Walker-Loud\cite{Walker-Loud:2023aa} categorized all results known at that time into two groups, (A) deeply bound dinucleons or (B) no bound state.
This classification shows that 
 the controversy is not caused by the difference of methods, the FV method or the HAL QCD method. Instead he pointed out that results with deeply bound dinucleons were obtained from FV spectra with  compact hexa-quark source operators\cite{Yamazaki:2012hi,NPLQCD:2012mex,Yamazaki:2015asa,Orginos:2015aya,Berkowitz:2015eaa} but results with no bound dinucleons comes from calculations with diffuse sources such as
wall sources (HAL QCD)\cite{Ishii:2012ssm,Inoue:2011ai}, distillation (Mainz)\cite{Francis:2018qch}, stochastic LapH (CoSMoN)\cite{Horz:2020zvv} or sparsenning momentum (NPLQCD)\cite{Amarasinghe:2021lqa}.

\begin{figure*}[htb]
\centering
%\vskip -2.5cm
  \includegraphics[height=4.4cm]{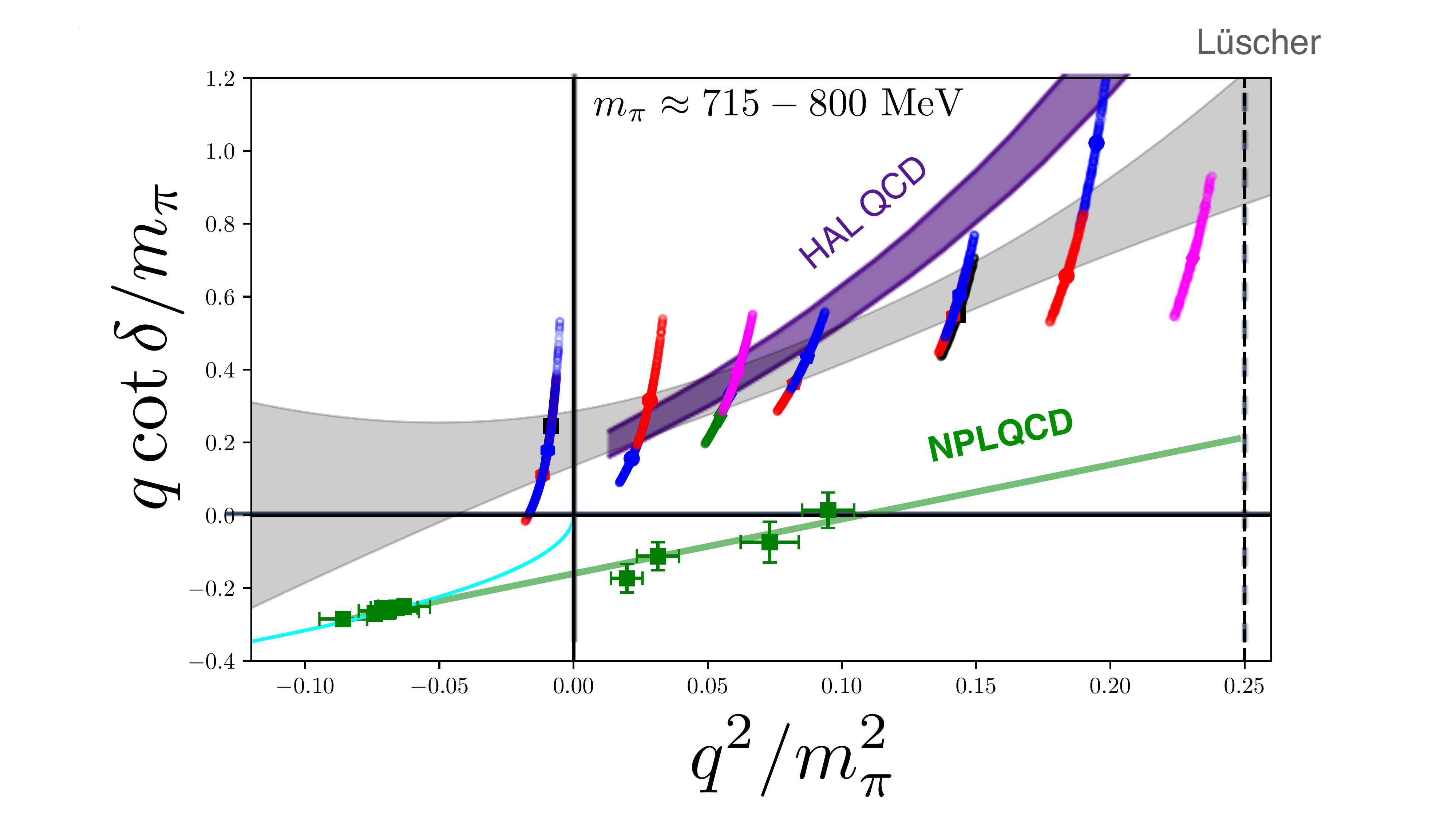}
  \includegraphics[height=4.8cm]{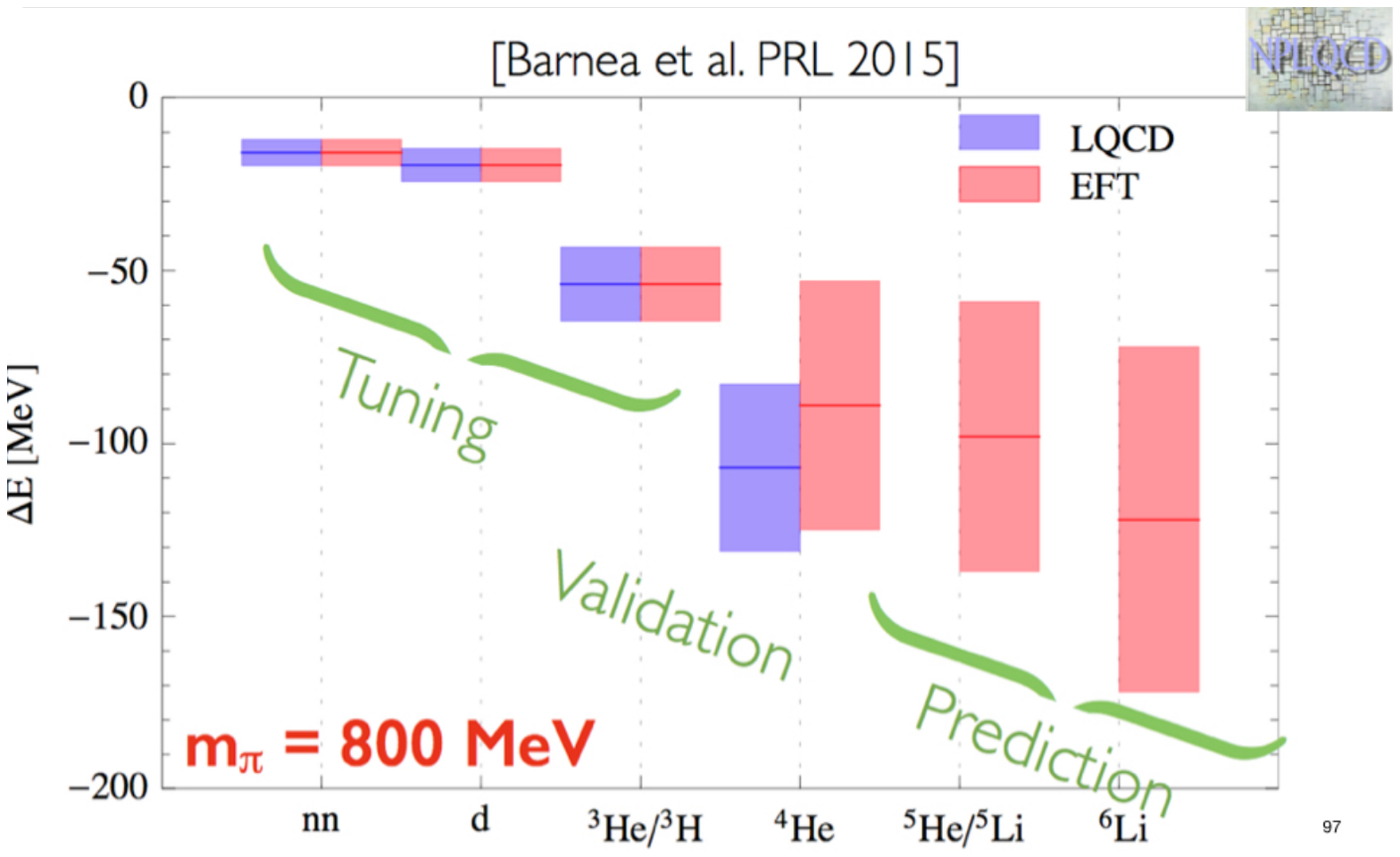}
 %\vskip -2.5cm
 \caption{ (Left) The $q\cot \delta(q)/m_\pi$ as a function of $q^2/m_\pi^2$ for the $NN$ scattering in the ${}^3S_1$ channel, obtained by the FV method with the diffused source at $m_\pi \simeq 715$ MeV (red, blue and magenta symbols) and  the hexa-quark source at $m_\pi \simeq  800$ MeV (green squares), together with the one by the HAL QCD method on the same configurations at $m_\pi \simeq 715$ MeV (purple band).
Figure taken from the slide by Nicholson at lattice 2022\cite{Nicholson:2022aa}.
(Right) A comparison of binding energies between the lattice QCD at $m_\pi=800$ MeV\cite{NPLQCD:2012mex} (blue) and the pionless EFT (red).
Deuteron, dineutron and ${}^3{\rm He}/{}^3{\rm H}$ are used to fix low energy constants of the pionless EFT,  while
 ${}^4{\rm He}$ is used to validate the fixed low energy constants. Then the pionless EFT predicts  binding energies of  ${}^5{\rm He}/{}^5{\rm Li}$
 and  ${}^6{\rm Li}$.  Figure taken from the slide by Beane at CD2015\cite{Beane:2015aa}.
}
% \vskip -0.5cm 
 \label{fig:Lat2022}
\end{figure*}

A comparative study on FV spectra between the wall source and the hexa-quark source has suggested that
the hexa-quark source tends to create a fake plateau for correlation functions of dinucleons, which can be misidentified as ``a deeply bound state''\cite{Iritani:2016jie,Iritani:2017rlk}.
The HAL QCD collaboration  performed a extensive study\cite{Iritani:2018vfn}
and found that a naive plateau identification employed by old studies in the FV method
cannot control systematics from excited state contaminations,
while the HAL QCD method does not suffer from this issue.
It was also explicitly shown that the plateaux from the hexa-quark source are indeed fake.
By addressing this issue using optimized (sink) operators, the HAL QCD collaboration showed
that the results from the FV method become consistent with those from the HAL QCD method, i.e., unbound NN systems.
New  calculations by the more sophisticated FV method followed.
At lattice 2022,  Nicholson\cite{Nicholson:2022aa} presented  Fig.~\ref{fig:Lat2022}(Left), which compares $q\cot \delta(q)/m_\pi$ as a function of $q^2/m_\pi^2$
among the HAL QCD method at  $m_\pi \simeq 715$ MeV (purple band) and  the FV method with the diffused source at $m_\pi \simeq 715$ MeV (red, blue and magenta symbols) and the hexa-quark source at $m_\pi \simeq  800$ MeV (green squares),  where $\delta(q)$ is the scattering phase shift of $NN({}^3S_1)$.
The first two are consistent with each other, showing the absence of the bound deuteron, while the last one is totally different in behaviors of $q\cot\delta (q)$, which intersects the bound state condition (cyan line) giving the deeply bound deuteron.\footnote{A combination of the FV formula and the effective range expansion of $q\cot\delta(q)$ was first introduced by the HAL QCD collaboration to check a normality of the bound stat spectra in the finite box\cite{Iritani:2017rlk} .} 
Now the community's consensus is that old FV results with hexa-quark sources have large uncertainties on the extraction of FV spectra, due to the fake plateau problem, as summarized in the statement by Walker-Loud at lattice 2023\cite{Walker-Loud:2023aa} that ``I believe the old results are wrong (including those I was involved with)''. 
Two nucleons are unbound at heavier pion masses.
Since  results claiming bound three or more  baryons also employed the hexa-quark sources\cite{Yamazaki:2012hi,NPLQCD:2012mex},
they  are  unreliable, too,  so are matrix element calculations of such ``bound states'' (for example, Re.~\cite{Winter:2017bfs}). 
See also  plenary talks by Morningstar and Green\cite{Green:2025rel}  in this conference (CD2024).

Could (chiral) effective theories invalidate  old FV data ?
Unfortunately, the answer is ``No''. 
Beane  made  Fig.~\ref{fig:Lat2022}(Right) at CD2015\cite{Beane:2015aa} using the data in Ref.~\cite{Barnea:2013uqa}, to claim
that  the pionless EFT, by fixing its low energy constants 
from  the old lattice result in the FV method\cite{NPLQCD:2012mex} on binding energies of deuteron, dineutron and ${}^3{\rm He}/{}^3{\rm H}$ at $m_\pi=800$ MeV,   
validates the lattice data on the binding energy of ${}^4{\rm He}$ at the same pion mass.
Similarly, the chiral perturbation theory  predicted pion mass dependences of  binding energy and momentum, 
which reasonably agree with the old lattice data in the FV method at four  pion masses above 300 MeV\cite{Baru:2015ira}. 
(See, however, the chiral EFT analysis which found inconsistency within old lattice data in the FV method\cite{Baru:2016evv}.)
It seems that the chiral effective theories can tell consistencies among lattice data but  cannot check correctness of  lattice data themselves.

In this talk, I will review recent results on interactions between two hadrons at almost physical pion mass, obtained exclusively by the HAL QCD method, emphasizing connections with chiral dynamics or experimental data.

\section{HAL QCD method}
We here briefly introduce the coupled channel HAL QCD method.
The coupled channel potential is extracted by the time-dependent HAL QCD method as 
\begin{eqnarray}
\left( {1+3\delta_c^2\over 8\mu_c} {\partial^2 \over \partial t^2} -{\partial \over \partial t} +{\nabla^2\over 2\mu_c}\right)
R^c{}_d (\vec r,t) =\sum_{c^\prime} \int d^3 r^\prime\, U^c{}_{c^\prime} (\vec r,\vec r^\prime) \Delta^c{}_{c^\prime} R^{c^\prime}{}_d(\vec r^\prime,t),
\end{eqnarray}
where $c,d$ represent channel indices, and 
the correlation function matrix  for two hadrons is defined by
 \begin{eqnarray}
 R^c{}_d (\vec r,t) &:=& \frac{\sum_{\vec x} \langle 0\vert H_{c_1}(\vec r +\vec x,t)   H_{c_1}(\vec x,t) \overline{\cal J}_d(0)\vert 0\rangle}{\sqrt{Z_{c_1}}\sqrt{Z_{c_2}} \exp\left[-(m_{c_1} + m_{c_2}) t\right]} .
 \end{eqnarray}
 Here $H_{c_1}$ and $H_{c_2}$ are local interpolating operators for hadrons while $ \overline{\cal J}_d(0)$ is a source operator creating two hadrons at  $t=0$, and $Z_{c_1}$ and $Z_{c_2}$ denote wave function renormalization factors for each hadron. 
 We also define the reduced mass $\mu_{c}=m_{c_1} m_{c_2}/(m_{c_1}+m_{c_2})$, 
 the size of mass difference $\delta_{c}= (m_{c_1}-m_{c_2})/(m_{c_1}+m_{c_2})$, 
 the factor $\Delta^c{}_d= \exp[ -(m_{d_1}+m_{d_2} - m_{c_1}-m_{c_2}) t]$, which compensates the threshold energy difference between $c$ and $d$ channels. 
 
In practice, the non-local potential $U^c{}_{c^\prime} $ is treated in the derivative expansion as
\begin{eqnarray}
U^c{}_{c^\prime} (\vec r,\vec r^\prime) =\left( V^c{}_{c^\prime} (\vec r) +\sum_{n=1} V^{(n) c}{}_{c^\prime} (\vec r) \nabla^n \right)\delta (\vec r-\vec r^\prime), 
\end{eqnarray}
and employ the leading order (LO) potential $V^c{}_{c^\prime} $ to obtain observables such as scattering phase shifts.
Systematic errors associated with the LO truncation of the derivative expansion   is estimated from $t$ dependences of potentials and physical observables,
which should be insensitive to sufficiently large $t$ if higher order contributions are small. 

All results in this talks have been obtained with gauges configurations on $96^4$ lattice at almost physical pion mass,
generated on the K-computer in Japan by the 2+1 flavor QCD with Iwasaki gauge action plus non-perturbatively  $O(a)$ improved
clover quark action at $a^{-1}\simeq 0.0846$ fm, so that  the spatial extension $La\simeq 8.1$ fm is larger enough\cite{Ishikawa:2015rho}.
 Some hadrons masses on this ensemble are $m_\pi \simeq146$ MeV, $m_K\simeq 525$ MeV, $m_N\simeq 955$ MeV, $m_\phi\simeq 1048$ MeV,
  $m_\Lambda\simeq 1140$ MeV,  $m_\Sigma\simeq 1222$ MeV, $m_\Xi \simeq 1355$ MeV and $m_\Omega\simeq 1712$ MeV.

\section{$\Lambda\Lambda$--$N\Xi$ interactions}
\subsection{Lattice results at almost physical pion mass}
\begin{figure*}[htb]
\centering
%\vskip -2.5cm
  \includegraphics[height=4.5cm]{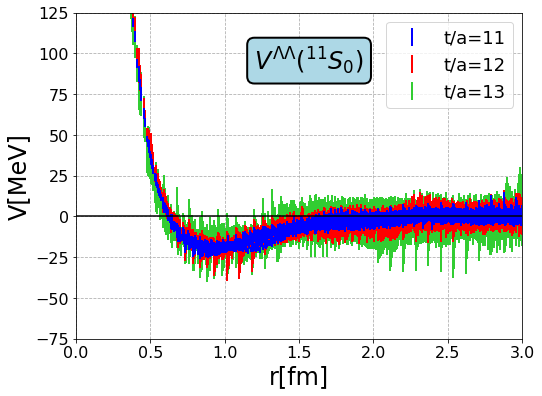}
   \includegraphics[height=4.5cm]{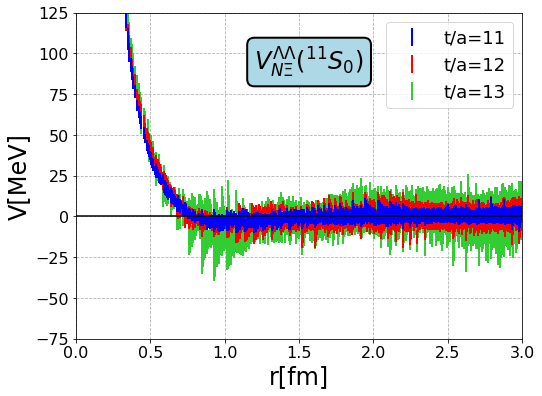}
     \includegraphics[height=4.5cm]{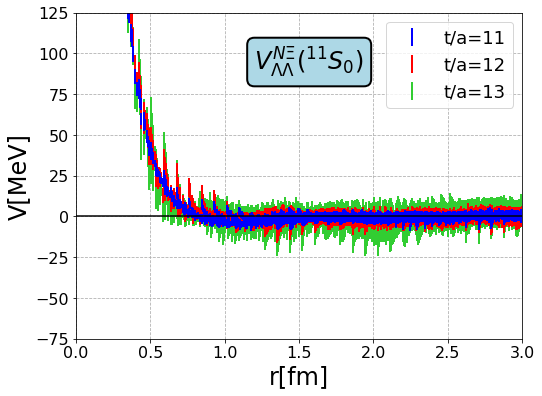}
        \includegraphics[height=4.5cm]{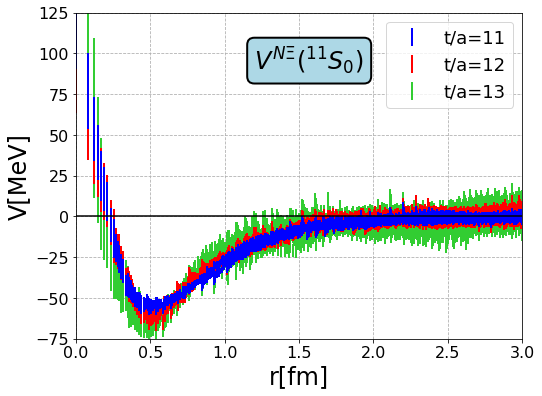}
 %\vskip -2.5cm
 \caption{ The $S$-wave $\Lambda\Lambda-N\Xi$ couple channel potentials in the ${}^{11}S_0$ as a function of the relative distance $r$.
 Here $V^{\Lambda\Lambda}$ (Upper-Left),  $V^{\Lambda\Lambda}_{N\Xi}$ (Upper-Right), $V_{\Lambda\Lambda}^{N\Xi}$  (Lower-Left) and $V^{N\Xi}$ (Lower-Right) stand for $\Lambda\Lambda\to \Lambda\Lambda$ (diagonal),  $N\Xi \to \Lambda\Lambda$ (off-diagonal), $\Lambda\Lambda\to N\Xi$ (off-diagonal) and   $N\Xi \to N\Xi$ (diagonal), respectively.  
 Figures taken from Ref.~\cite{HALQCD:2019wsz}.}
% \vskip -0.5cm 
 \label{fig:Potential-00S0}
\end{figure*}

In previous studies by the HAL QCD collaboration\cite{Inoue:2010es,Inoue:2011ai,Inoue:2010hs}, the H dibaryon ($uuddss$ state) has appeared as a bound state at heavy pion mass in the flavor SU(3) limit ($m_u=m_d=m_s$). 
It has been observed that the binding energy decreases  while a size of the bound state increases as the pion mass decreases.
What happens in real world, where the pion mass becomes much lighter and the flavor SU(3) is violated ?
Three thresholds appear in the H dibaryon channel at $2m_\Lambda\simeq$ 2280 MeV, $m_N + m_\Xi \simeq$ 2311 MeV and $ 2m_\Sigma \simeq$ 2444 MeV.
We have  calculated the $\Lambda\Lambda-N\Xi$ coupled channel interaction at  almost physical pion mass,
ignoring the $\Sigma\Sigma$ channel as its threshold is much higher.

In Fig.~\ref{fig:Potential-00S0}, we present the $S$-wave $\Lambda\Lambda - N\Xi$ coupled channel potentials in the ${}^{11}S_0$ channel obtained at almost physical pion mass,
where we use the notation ${}^{2I+1,2s+1}S_{J}$ to specify the total isospin $I$, total spin $s$ and the total angular momentum $J$ of $S$-wave states.
First of all, $t$ dependences of potentials are small, showing that the LO truncation of the derivative expansion is reasonably good.
Off-diagonal potentials are non-zero only at short distance,  indicating that the channel mixing between $\Lambda\Lambda$ and $N\Xi$ is negligible
at low energies.
Within diagonal potentials, the attraction in the $N\Xi$ channel is much stronger than $\Lambda\Lambda$.
This tendency was observed already at heavier pion masses.

\begin{figure*}[htb]
\centering
%\vskip -2.5cm
  \includegraphics[height=4.5cm]{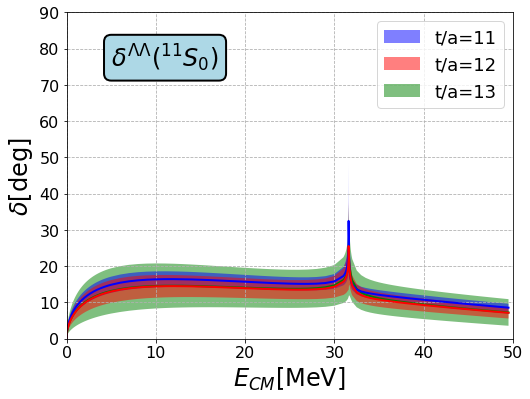}
   \includegraphics[height=4.5cm]{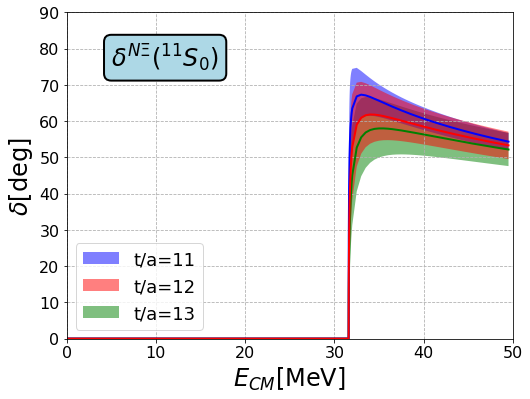}
 %\vskip -2.5cm
 \caption{(Left) $\Lambda\Lambda$ scattering phase shift and (Right) $N\Xi$ scattering phase shift  in the ${}^{11}S_0$ channel as a function of the center of mass energy $E_{CM}$ from the $\Lambda\Lambda$ threshold.   Figures taken from Ref.~\cite{HALQCD:2019wsz}. }
% \vskip -0.5cm 
 \label{fig:Phase_Shift-00S0}
\end{figure*}
Fig.~\ref{fig:Phase_Shift-00S0} shows scattering phase shifts of $\Lambda\Lambda$ (Left) and $N\Xi$ (Right) as a a function of the center of mass energy $E_{CM}$ from the $\Lambda\Lambda$ threshold\cite{HALQCD:2019wsz}.
A positive but small $\Lambda\Lambda$ scattering phase shift indicates that neither bound state nor resonance exits in this channel 
because of weak attraction in the diagonal  $\Lambda\Lambda$ potential.
In particular, no H dibaryon appears near the  $\Lambda\Lambda$ threshold and we obtain $a_0^{\Lambda\Lambda}=0.8(3)$ fm for the $\Lambda\Lambda$ scattering length.
We observe a sharpe enhancement and rapid drop of the  $\Lambda\Lambda$ scattering phase shift near the $N\Xi$ threshold,
which is caused by the off-diagonal potential 
between $\Lambda\Lambda$ and $ N\Xi$. 
The $N\Xi$ scattering, on the other hand, exhibits a sharpe increase  near the $N\Xi$ threshold due to the significant attraction of the diagonal 
$N\Xi$ potential, whose maximum almost reaches to 60 degrees.
An analysis of the coupled channel scattering S-matrix in complex energy planes reveals that   H dibaryon appears as  a nearly virtual  state 
with very small width near the $N\Xi$ threshold at $m_\pi\simeq 146$ MeV and $a\simeq  0.0846$ fm.

\subsection{Comparison with experiments}
RHIC and  LHC have measured  two hadron correlations, 
$C_{AB} (Q) = N_{AB}^{\rm pair}(Q) / N_A(Q) N_B(Q)$,
where $N_{AB}(Q)$ is a number of pairs of hadron $A$ and hadron $B$ in the final states of heavy ion collisions at energy $Q$,
while $N_A(Q)$ and $N_B(Q)$ are those of the single hadron. 
One can estimate $N_{AB}(Q)$ as
\begin{eqnarray}
N_{AB}(Q) = \int {d^3 p_A \over E_A} {d^3 p_B \over E_B} N_{AB}(\vec p_A,\vec p_B) \delta\left(Q -\sqrt{-q^2}\right), \quad q^2=q_\mu q^\mu,
\end{eqnarray}
where $q^\mu$ is the relative 4 momentum, and 
\begin{eqnarray}
N_{AB}(\vec p_A,\vec p_B) &\simeq&  \int d^4x d^4y\, S_{AB}(x,\vec p_A) S_B(y,\vec p_B) \vert \Psi(x,y, \vec p_A,\vec p_B)\vert^2 .
\end{eqnarray}
 Here $S_A$ and $S_B$ are source functions of each hadron and   $\Psi(x,y, \vec p_A,\vec p_B)$ is the 2-body wave function between $A$ and $B$, obtained from a solution to the Schr\"odinger equation with a given potential.
 Therefore, if source functions are approximately known, one can test hadron interactions against two hadron correlations obtained in experimental
 through the above formula, and vice versa. This type of analysis for hadron interactions through heavy ion collision experiments is called ``Femtoscopy''\cite{ALICE:2019hdt}.

 \begin{figure*}[htb]
\centering
%\vskip -2.5cm
  \includegraphics[height=9cm]{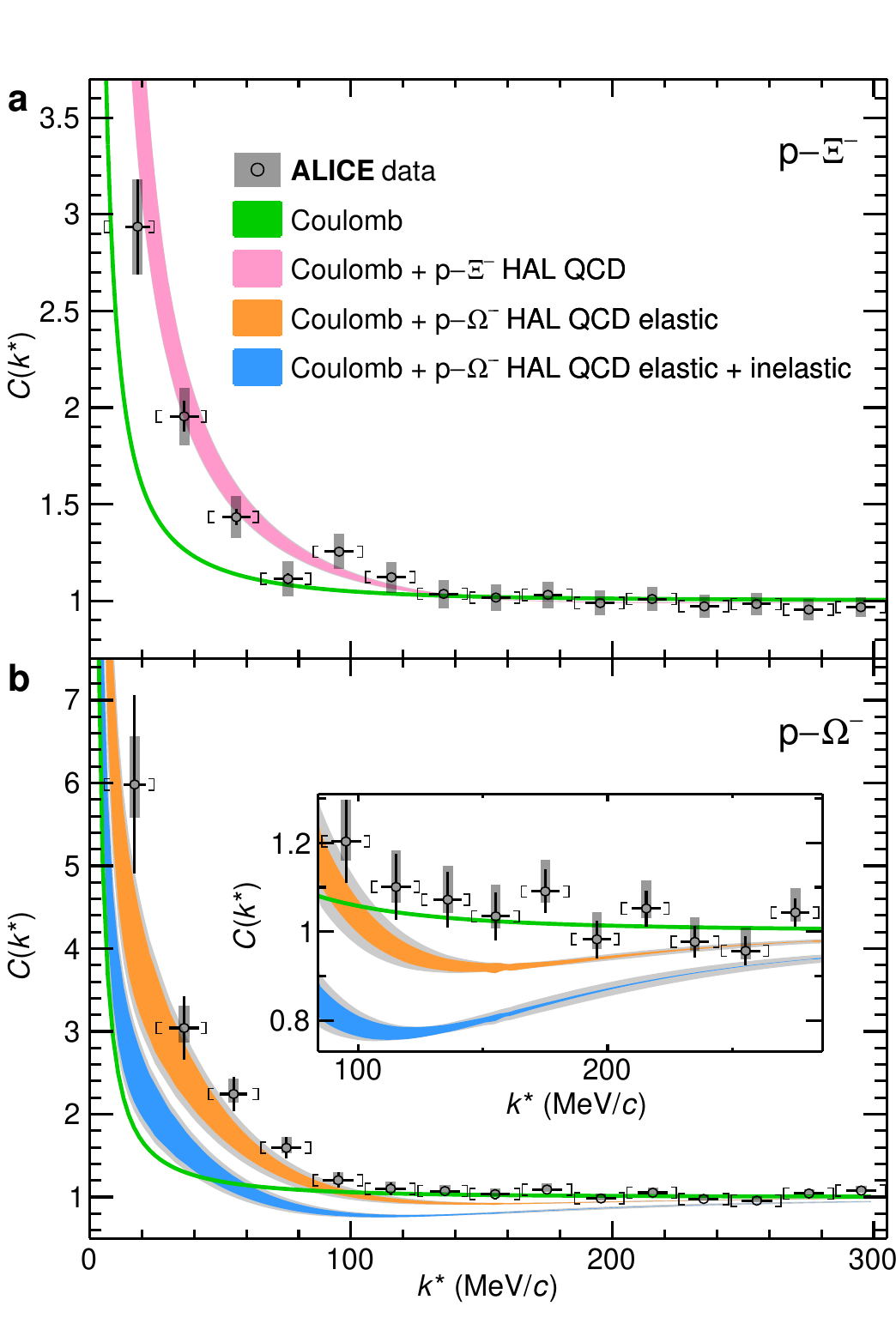}
% \vskip -2.5cm
 \caption{(Upper) Correlations between $p$ (proton) and $\Xi^-$  as a function of $k^*=\vert \vec k^*\vert$
with the relative momentum $\vec k^* =\vec p_A-\vec p_B$, where experimental data\cite{ALICE:2020mfd} are shown as black symbols while
theoretical predictions by Coulomb with or without  strong HALQCD interactions\cite{HALQCD:2019wsz}  are pink or green bands, respectively.
 (Lower) Correlations between $p$ and $\Omega^-$,  where experimental data\cite{ALICE:2020mfd} are shown as black symbols while
theoretical predictions by Coulomb,  including HALQCD elastic interactions\cite{HALQCD:2018qyu} or  including both elastic \& inelastic interactions are green, orange or blue bands, respectively.
 Figures taken from Ref.~\cite{ALICE:2020mfd}. }
 %\vskip -0.5cm 
 \label{fig:Femto}
\end{figure*}
Fig.~\ref{fig:Femto} (Upper)  shows a comparison of hadron correlations between $p$ (proton) and $\Xi^-$,
where black symbols represent experimental data obtained at LHC by  ALICE Collaboration \cite{ALICE:2020mfd}
while pink and green bands give theoretical predictions by Coulomb  with and without  HALQCD interactions, respectively. As seen in the figure, the correlation obtained by  the Coulomb with HALQCD potential agrees  with the experimental data much better than the one by the Coulomb potential alone.  

\vskip -0.5cm
\section{$N\Omega$ dibaryon}
\begin{figure*}[htb]
\centering
%\vskip -2.5cm
  \includegraphics[height=4.5cm]{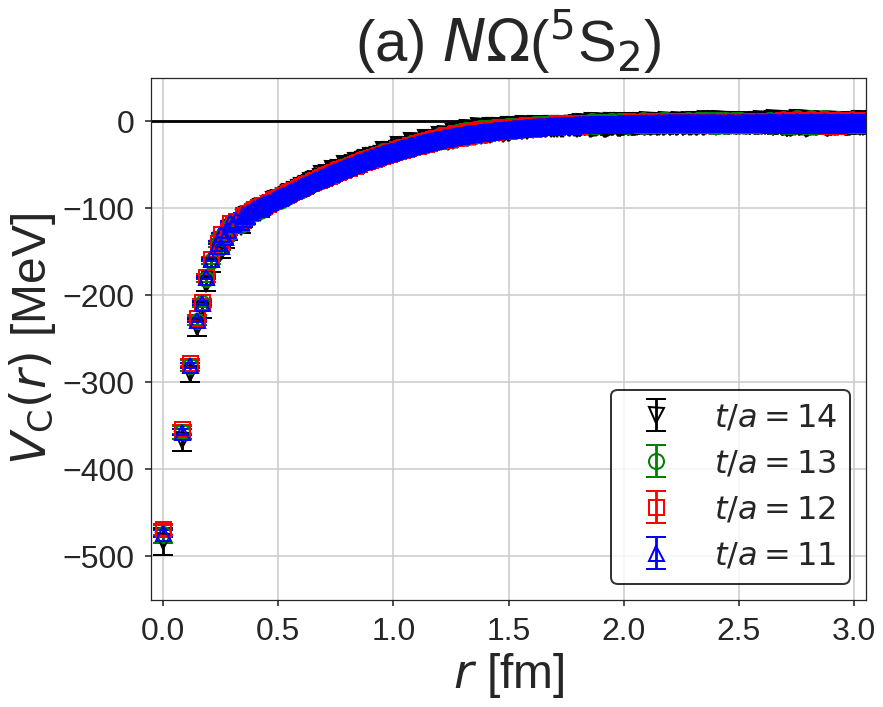}
   \includegraphics[height=4.5cm]{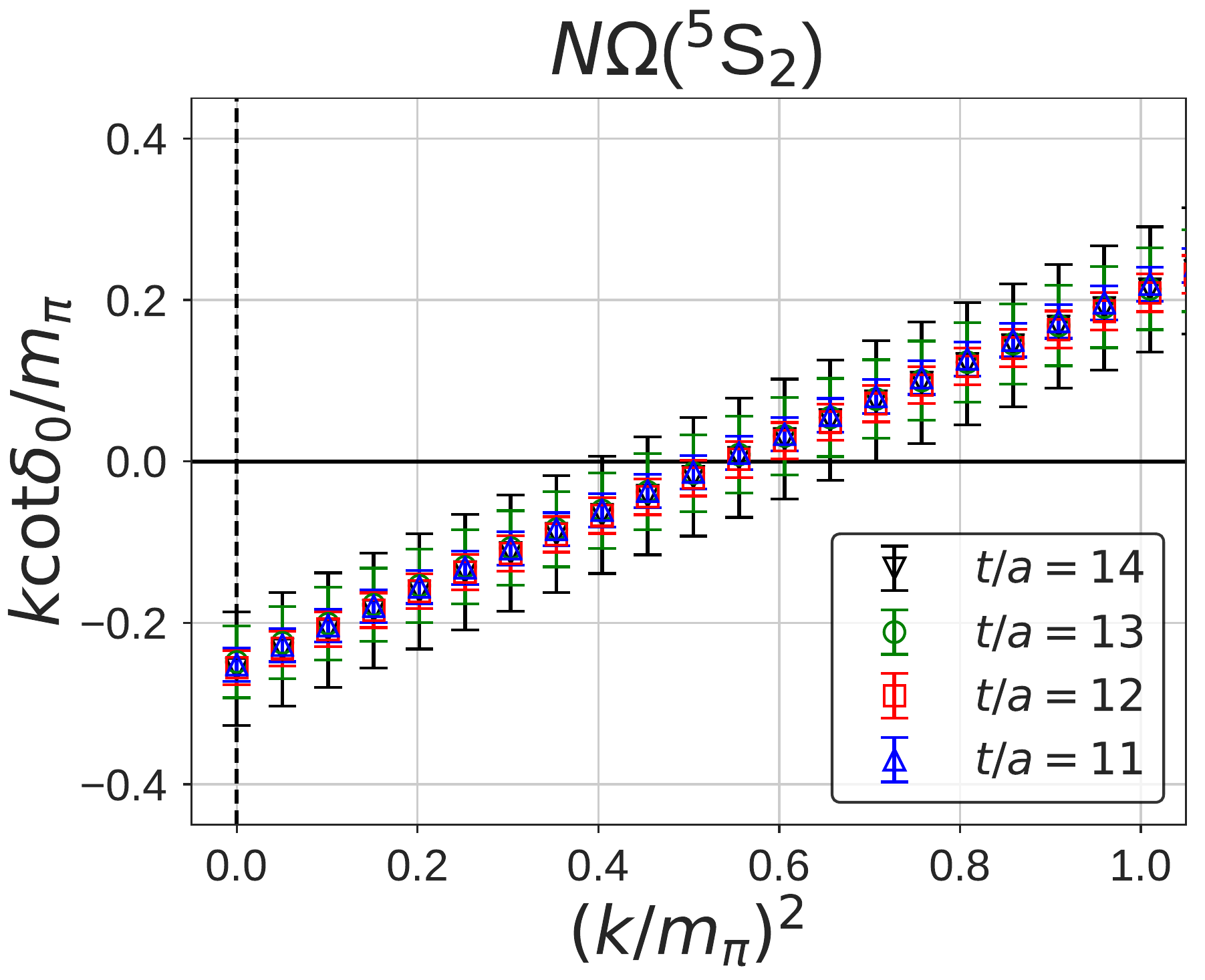}
 %\vskip -2.5cm
 \caption{(Left) The $N\Omega$ potential in ${}^5S_2$ channel at $t/a=11$ (blue up triangles), 12 (red squares), 13 (green circles) and 14 (black down triangle). (Right) The $k\cot\delta_0(k)/m_\pi$ as a function of  $(k/m_\pi)^2$, where $\delta_0(k)$ is the $S$-wave scattering phase shift  of the $N\Omega({}^5S_2)$ system. Figures taken from Ref.~\cite{HALQCD:2018qyu}. }
 %\vskip -0.5 cm 
 \label{fig:Pot_NOmega}
\end{figure*}
The $N\Omega$ potential in the ${}^5S_2$ channel ($s=2$ and $J=2$) has been calculated at almost physical pion mass, as shown in Fig.~\ref{fig:Pot_NOmega} (Left), which is attractive at all distances without  repulsive core.  
This behavior is qualitatively the same at $m_\pi\simeq 875$ MeV\cite{HALQCD:2014okw}.
At $m_\pi\simeq 146$ MeV, there are two thresholds,  $\Lambda\Xi ({}^{31}D_2)$ and  $\Sigma\Xi ({}^{31}D_2)$, below the $N\Omega({}^5S_2)$ threshold $E_{\rm th}^{N\Omega} \simeq 2666$ MeV.  Explicitly $E_{\rm th}^{\Lambda\Xi ({}^{31}D_2)} \simeq 2495$ (2514) MeV and  $E_{\rm th}^{\Sigma\Xi ({}^{31}D_2)} \simeq 2577$ (2595) MeV  at $L=\infty$ (8.1 fm) with $L$ being the spacial extension.
However only the single channel analysis  was made in Ref.~\cite{HALQCD:2014okw} by assuming small couplings to $D$-wave states,
which  is supported by the weak $t$ dependence of the potentials.
In future,  the coupled channel analysis will be needed to check the current result.

Fig.~\ref{fig:Pot_NOmega} (Right) gives the $k\cot \delta_0(k)$ with $\delta_0(k)$ being the $S$-wave scattering phase shift, whose 
ERE fit
leads to the scattering length $a_0=-5.30(0.44)(^{+0.16}_{-0.01})$ fm and the effective range $r_{\rm eff} = 1.60(0.01)(^{+0.02}_{-0.01})$ fm, where
the central value and the statistical errors are estimated at $t/a=12$ while the systematic errors in the last parentheses are estimated from the largest differences of the central values ($t/a=12$) among  $t/a=11,13,14$.  
The behavior of the  $k\cot \delta_0(k)$ suggests an existence of a bound state in this channel.
A solution of the Schr\"odinger equation with the potential %in Fig.~\ref{fig:Pot_NOmega} (Left) 
gives
the binding energy $B=1.54(0.30)(^{+0.04}_{-0.10})$ MeV and the root mean square distance $\sqrt{\langle r^2\rangle} = 3.77(0.31)(^{+0.11}_{-0.01})$ fm.
This binding energy is much smaller than $B= 18.9(5.0)(^{+12.1}_{-1.8})$ MeV obtained at $m_\pi\simeq 875$ MeV\cite{HALQCD:2014okw}.
The smallness of the binding energy at $m_\pi\simeq146$ MeV is caused by the short range nature of the potential, though it is attractive everywhere.
As the pion mass decreases, smaller masses of $N$ and $\Omega$ increase the kinetic energy of the system, so that it is loosely bound  like the deuteron and the root mean square distance becomes larger.
If  an extra attraction in the $p\Omega^-$ system due to the Coulomb interaction is included by adding $-\alpha/r$ term to the potential with $\alpha:=e^2/(4\pi)=1/137.036$, we obtain
$B_{p\Omega^-}=2.46(0.34)(^{+0.04}_{-0.11})$ MeV and $\sqrt{\langle r^2\rangle_{p\Omega^-}} = 3.24(0.19)(^{+0.06}_{-0.00})$ fm.

Fig.~\ref{fig:Femto} (Lower) shows a comparison of hadron correlations between $p$ and $\Omega^-$,
where black symbols represent experimental data obtained by  ALICE Collaboration\cite{ALICE:2020mfd}
while green,  orange or blue bands give theoretical predictions by Coulomb only,  including HALQCD elastic in Fig.~\ref{fig:Pot_NOmega} or further adding inelastic interactions, respectively. 
At smaller $k^*$, the correlation obtained by  the Coulomb + HALQCD elastic agrees  with the experimental data much better than the other two,
while the Coulomb alone works better around $k^* \simeq  100 - 200$  MeV.  
Since the inclusion of  inelastic contributions estimated by some model makes agreement worse,
it will be important to evaluate  inelastic contributions directly by the couple channel analysis in the HAL QCD method.  

\section{Tetra-quark state $T_{cc}$}
\begin{figure*}[htb]
\centering
%\vskip -2.5cm
  \includegraphics[height=4.5cm]{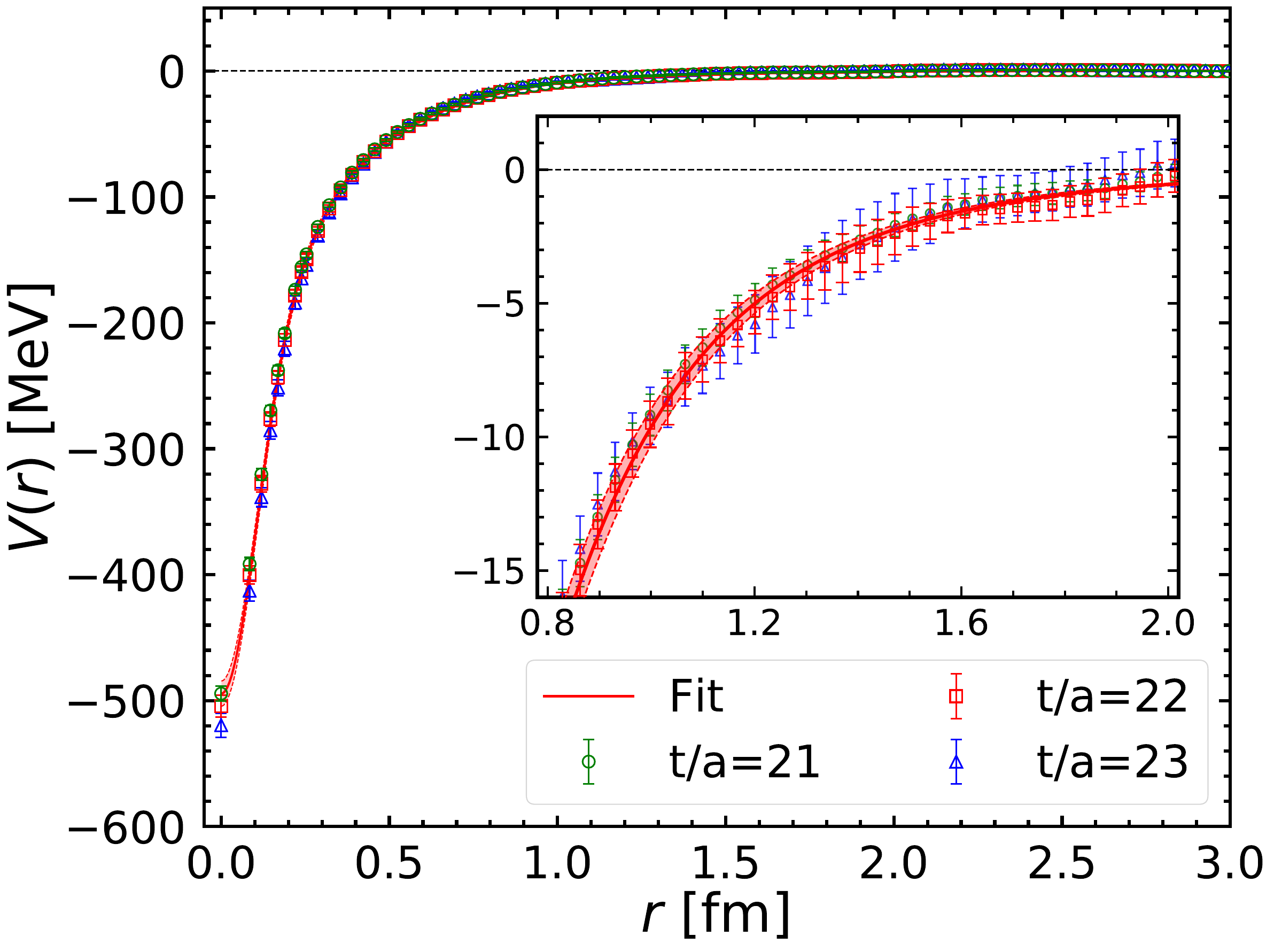}
   \includegraphics[height=4.5cm]{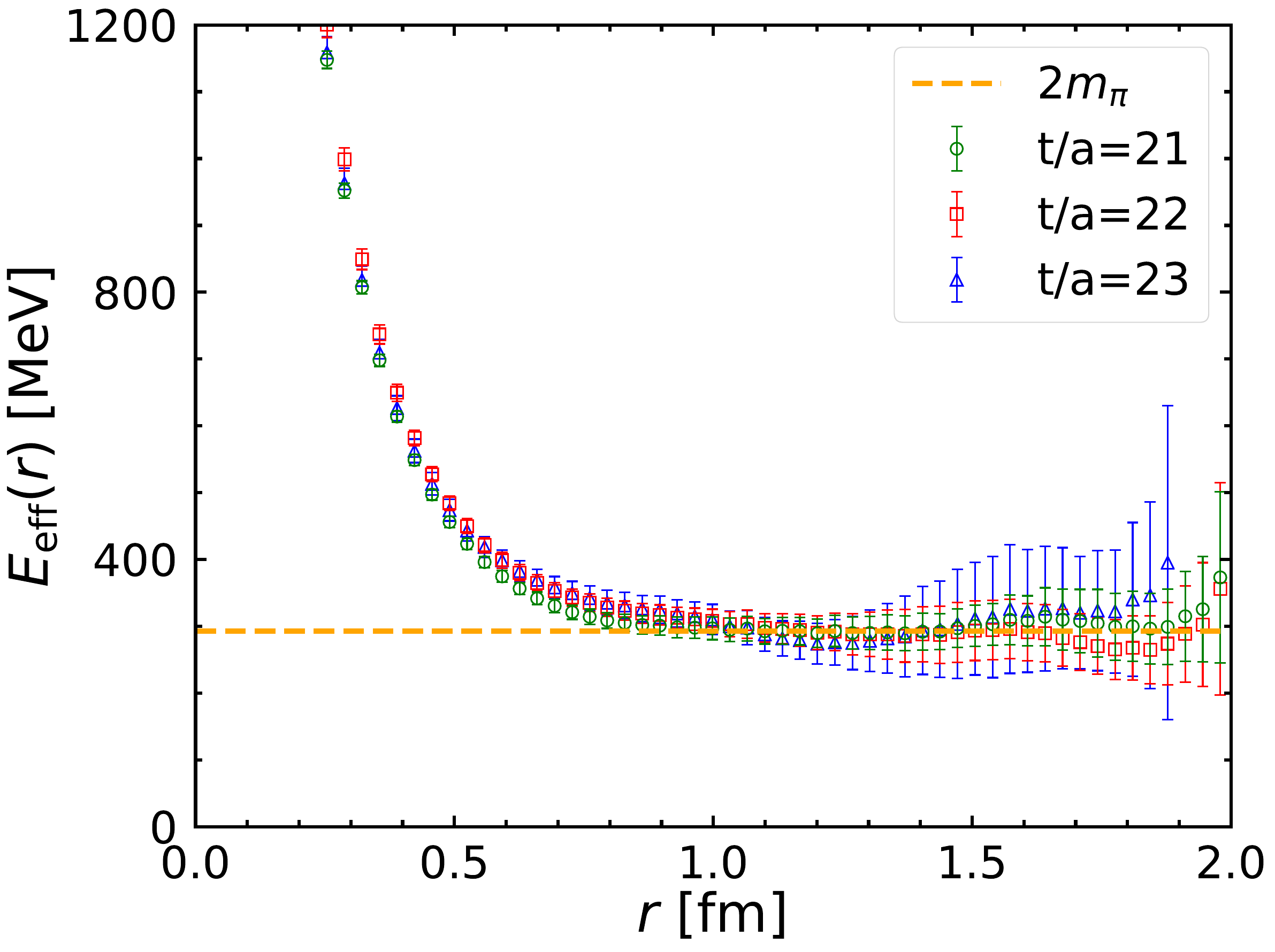}
 %\vskip -2.5cm
 \caption{(Left) The $D^*D$ potential $V(r)$ in the $I=0$ and $S$-wave channel at $t/a= 21$ (green circles), 22 (red squares) and 23 (blue triangles),
 while the red band shows the fitted potential with $V_{\rm fit} (r; m_\pi)$  for $t/a=22$.
 (Right) The effective energy $E_{\rm eff}(r)$ in space as a function of $r$ at three $t/a$,
  while the orange dashed line denotes $2m_\pi$ with $m_\pi = 146.4$ MeV.
 Figures taken from Ref.~\cite{Lyu:2023xro}. }
% \vskip -0.5cm 
 \label{fig:pot_D*D}
\end{figure*}

A tetra-quark state which contains two heavy (anti-)quarks is  a genuine tetra-quark state, since it doesn't couple to ordinary mesons in the the strong interaction due to an absence of pair annihilation channels.
Experimentally, LHCb Collaboration reported the observation of the tetra-quark state $T_{cc}(cc\bar u\bar d)$, which appears 360 keV below the $D^{*+} D^0$ thresholds 
having    $(I,J^P)=(0,1^+)$\cite{LHCb:2021vvq,LHCb:2021auc}.
Since the scattering length in the $D^* D$ channel in lattice QCD at heavier pion masses shows the significant pion mass dependence, as seen in Fig.~1 of Ref.~\cite{Lyu:2023xro}, the  calculation at almost physical pion mass is called for.

There are two channels, $D^0\pi^+ D^0$ (3869.25 MeV) and $D^+\pi^0 D^0$ (3869.48 MeV), below the
 $D^{*+} D^0$ threshold (3875.10 MeV) or $D^{*0} D^+$ threshold (3876.51 MeV) in Nature, 
 while, in the isospin symmetric 2+1 flavor lattice QCD at slightly heavier pion mass $m_\pi =146.4$ MeV,
 $D^*D$ threshold (3869.3 MeV) is smaller than $D\pi D$ threshold at 3902.8 (3974.3 ) MeV on $L\to\infty$  ($L\simeq 8.1$ fm),
  %or 3974.3 MeV on $L\simeq 8.1$ fm,  
  so that  the single  $D^*D$ channel analysis is justified.

Fig.~\ref{fig:pot_D*D} (Left) presents the $D^*D$ potential as a function of $r$, which is attractive at all distances with a long range tail. 
It turned out that the potential is consistent with (Yukawa)$^2$ at large $r$ and is fitted by two Gaussians plus  (Yukawa)$^2$,
\begin{eqnarray}
V_{\rm fit} (r; m_\pi) &=& \sum_{i=1,2} a_i e^{-(r/b_i)^2} + a_3 \left( 1-  e^{-(r/b_3)^2} \right)^2 \left( {e^{-m_\pi r} \over r}\right)^2,
\label{eq:fit_Tcc}
\end{eqnarray}
as shown by a red band in the figure. 
Fig.~\ref{fig:pot_D*D} (Right) gives an effective energy in space,  defined by
\begin{eqnarray}
E_{\rm eff}(r) &=& - \frac{\ln [ V(r) r^2/a_3 ]}{r}
\end{eqnarray}
  as a function of $r$, where $a_3$ is taken from the fitted value in \eqref{eq:fit_Tcc}.
  Since $E_{\rm eff}(r)$ converges to $2m_\pi$ at large $r$, the 2-pion rather than the 1-pion contribution dominate at long distance.
  Thus the 1-pion exchange is not a dominant contribution for $T_{cc}$.

\begin{figure*}[tbh]
\centering
%\vskip -2.5cm
  \includegraphics[height=4.5cm]{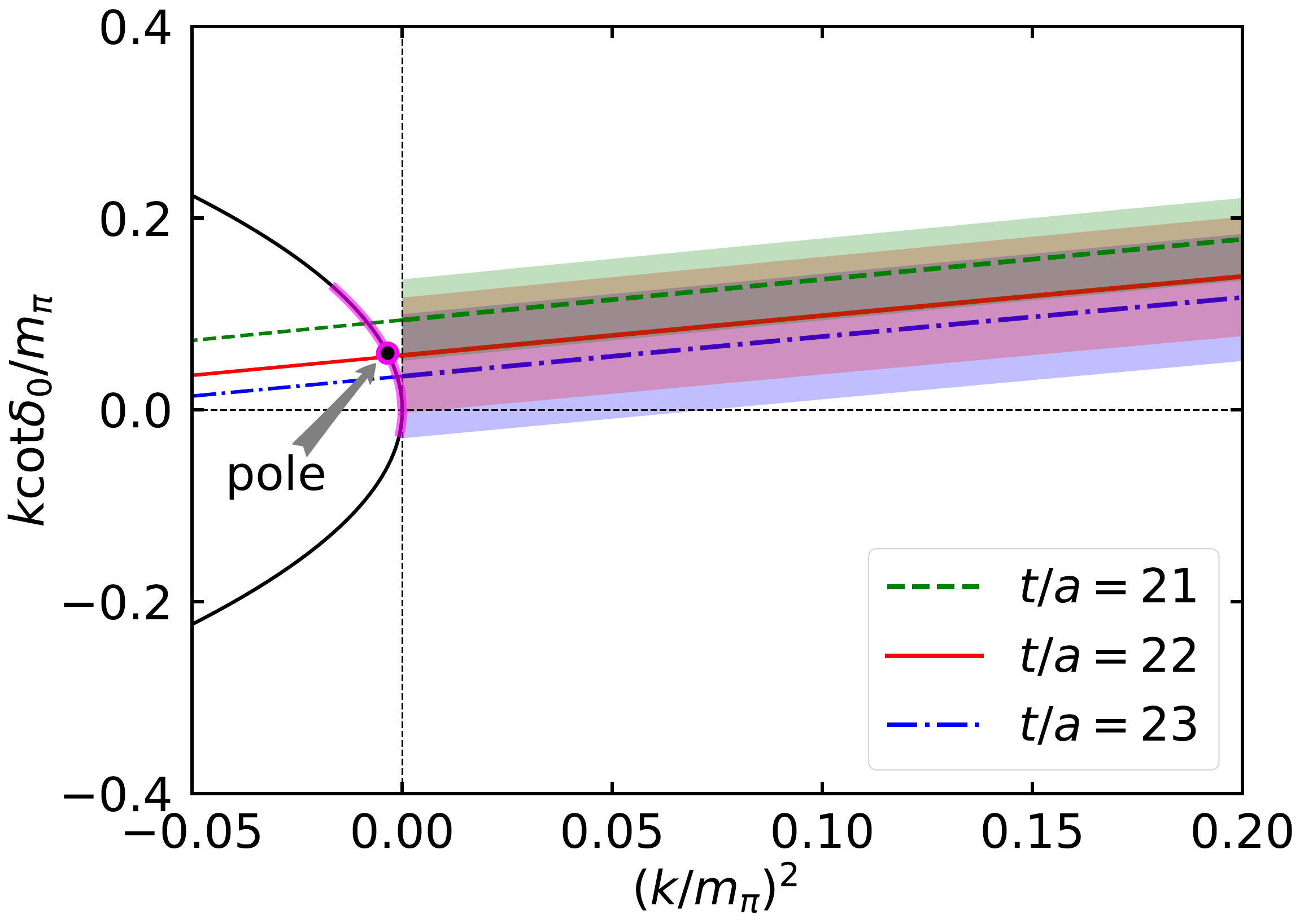}
   \includegraphics[height=4.5cm]{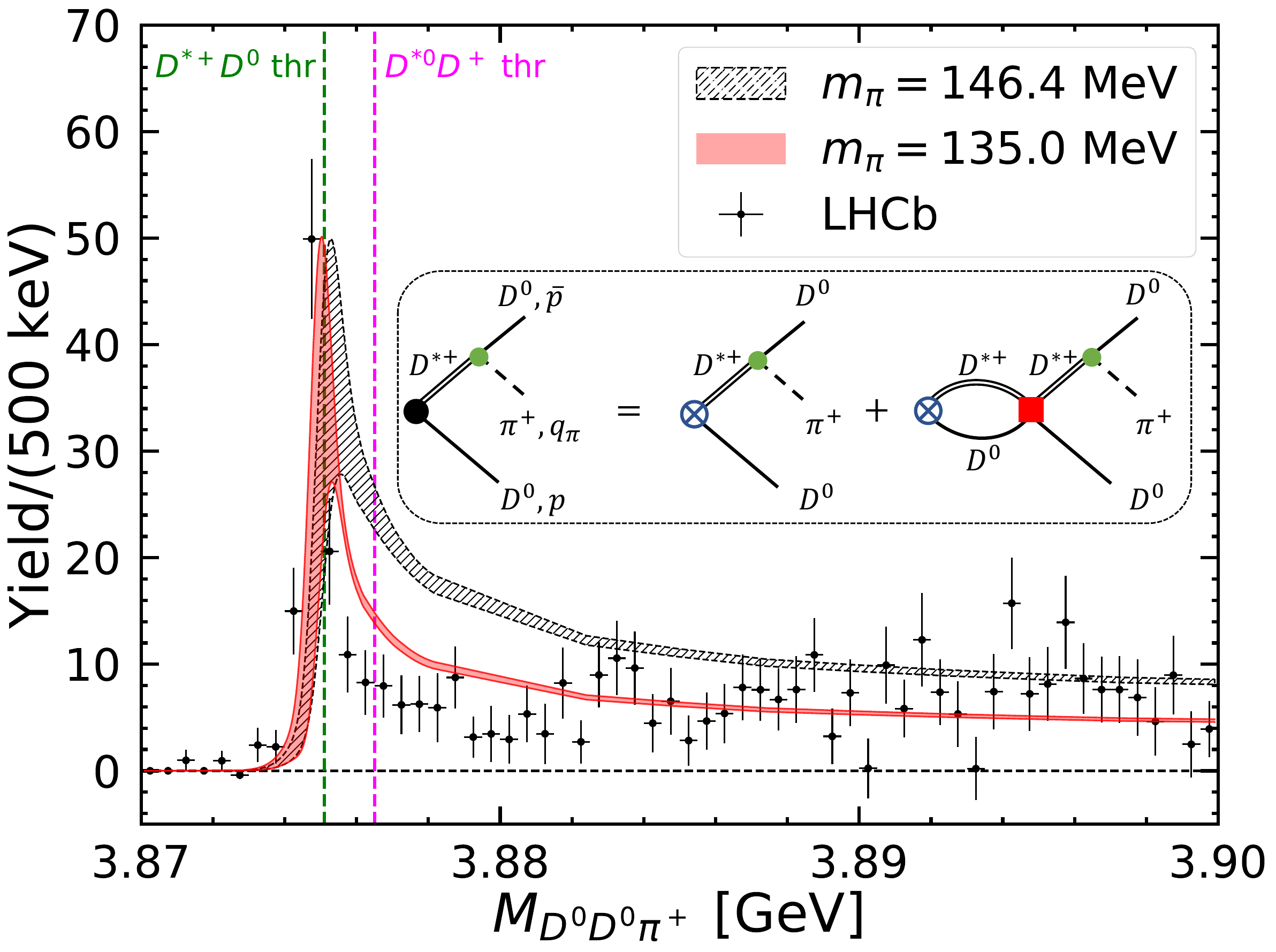}
 %\vskip -2.5cm
 \caption{(Left) The $k\cot \delta_0(k)/m_\pi$ %as a function of $(k/m_\pi)^2$ 
 for the $D^*D$ scattering 
 in the $I=0$ and $S$-wave channel.
 The intersection of the $k\cot \delta_0(k)/m_\pi$ and the black solid line ($\pm\sqrt{-(k/m_\pi)^2}$) denotes
 the pole of the scattering amplitude.
 (Right) The $D^0D^0\pi^+$ mass spectrum for theoretical results with $V_{\rm fit}(r;m_\pi)$ 
 at $m_\pi=146.4$ MeV (black band) and $m_\pi=135.0$ MeV (red band) and
 experimental data by LHCb (black points).
 Figures taken from Ref.~\cite{Lyu:2023xro}. }
 %\vskip -0.5cm 
 \label{fig:kcotd_Tcc}
\end{figure*}
Fig.~\ref{fig:kcotd_Tcc} (Left) shows $k\cot\delta_0(k)/m_\pi$ as a function of $(k/m_\pi)^2$, where
the phases shift $\delta_0(k)$ for the $D^*D$ scattering in the $I=0$ and $S$-wave channel
are calculated with the fitted potential $V_{\rm fitt}$.
This analysis gives
an inverse of the scattering length $a_0$ as $a_0^{-1}$ [fm$^{-1}$] $= 0.05 (5)(^{+2}_{-2}) $.
Since the $k\cot\delta_0(k)/m_\pi$ intersects the $+\sqrt{-(k/m_\pi)^2}$ (black solid line), as denoted by ``pole'' in the figure
with magenta line being total errors,
there appears one shallow virtual state (not bound state) at $k=i\kappa_{\rm pole}$ with $\kappa_{\rm pole}= -8(8)(^{+3}_{-5})$ MeV,
equivalently  $E_{\rm pole}=  -59(^{+53}_{-99})(^{+2}_{-67})$ keV.
This virtual pole exists above the left-hand cut for the 1-pion exchange, which appears at $k^2/m_\pi^2\le -0.02$ in this lattice setup.
The issue on the left-hand cut for $T_{cc}$ and other cases has been discussed in lattice 2024 by the author\cite{Aoki:2025jvi}.
See also Refs.~\cite{Du:2023hlu,Meng:2023bmz,Collins:2024sfi} for the issue.

\begin{table}[htp]
\caption{The inverse of the scattering length  $1/a_0$, the effective range $r_{\rm eff}$, the pole position $\kappa_{\rm pole}$ and 
$E_{\rm pole}$ at $m_\pi= 146.4$ MeV and 135.0 MeV.
}
\begin{center}
\vskip -0.5cm
\begin{tabular}{|c|llll|}
\hline\hline
$m_\pi$ (MeV) & $1/a_0$[fm$^{-1}$] & $r_{\rm eff}$ [fm] & $\kappa_{\rm pole}$ [MeV] &  $E_{\rm pole}$ [keV] \\
\hline
146.4 &  0.05(5)($^{+2}_{-2}$) &1.12(3)($^{+3}_{-8}$)& -8(8)($^{+3}_{-5}$) & -59($^{+53}_{-99}$)($^{+2}_{-67}$)\\
135.0 & -0.03(4)  &1.12(3) & 5(8) & -45($^{+41}_{-78}$)\\
\hline 
\end{tabular}
\end{center}
\label{tab:comp}
\vskip -0.3cm
\end{table}%
We estimate  an impact of the small difference in the pion mass on observables  between our simulation and Nature, 
by replacing $m_\pi$ in $V_{\rm eff}(r;m_\pi)$ from 146.4 MeV to 135.0 MeV to calculate scattering observables.
Results are summarized in Table~\ref{tab:comp}.
As seen from signs of $1/a_0$ and $\kappa_{\rm pole}$,
$T_{cc}$ moves from a shallow virtual state to a shallow bound state
by the extrapolation from $m_\pi=146.4$ MeV to $m_\pi = 135.0$ MeV in terms of the potential.
Alternatively, an extrapolation of  $1/a_0$ linear in $m_\pi^2$ gives $1/a_0= -0.01(9)$ [fm$^{-1}$] at $m_\pi=135.0$ MeV\cite{Lyu:2023xro}.
Thus two extrapolations, $1/a_0$[fm$^{-1}$]= -0.03(4) and -0.01(9), are consistent within errors.
  
Fig.~\ref{fig:kcotd_Tcc} (Right) shows the $D^0D^0\pi^+$  mass spectrum, where black points denote  LHCb data while
black and red bands represent mass spectra  based on $V_{\rm fit}(r;m_\pi)$ with $m_\pi=146.4$ MeV and 135.0 MeV, respectively\cite{Lyu:2023xro}.
The potential $V_{\rm fit}$ was employed to estimate the rescattering between $D^{*+}$ and $D^0$, denoted by the red circle in the third diagram of the insect in the figure.
It seems that the potential extrapolated to  $m_\pi = 135$ MeV explains LHCb data better. 

\section{$N$--$\phi$ interaction and the 2-pion tail}
\begin{figure*}[htb]
\centering
%\vskip -2.5cm
  \includegraphics[height=4.5cm]{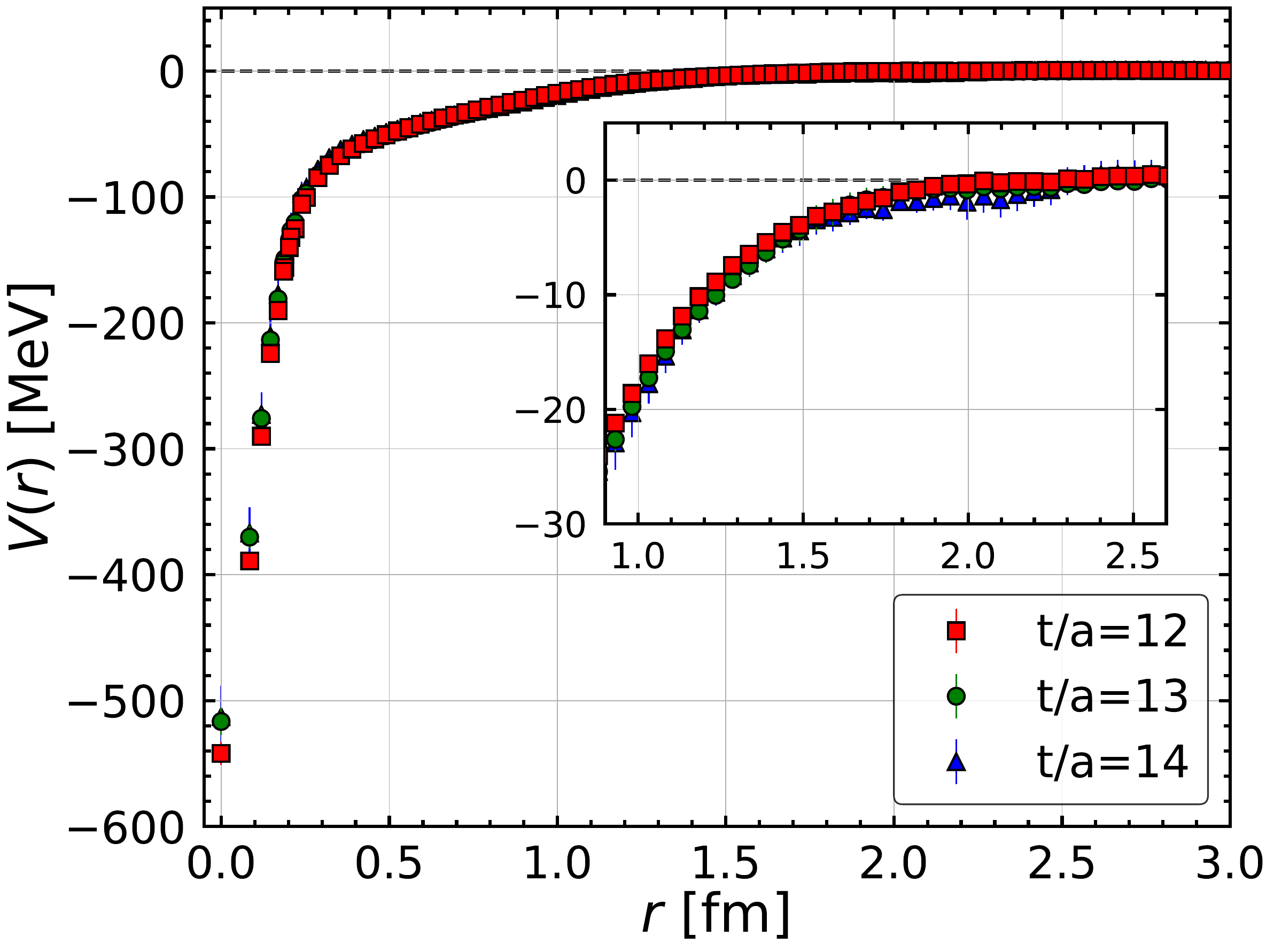}
   \includegraphics[height=4.5cm]{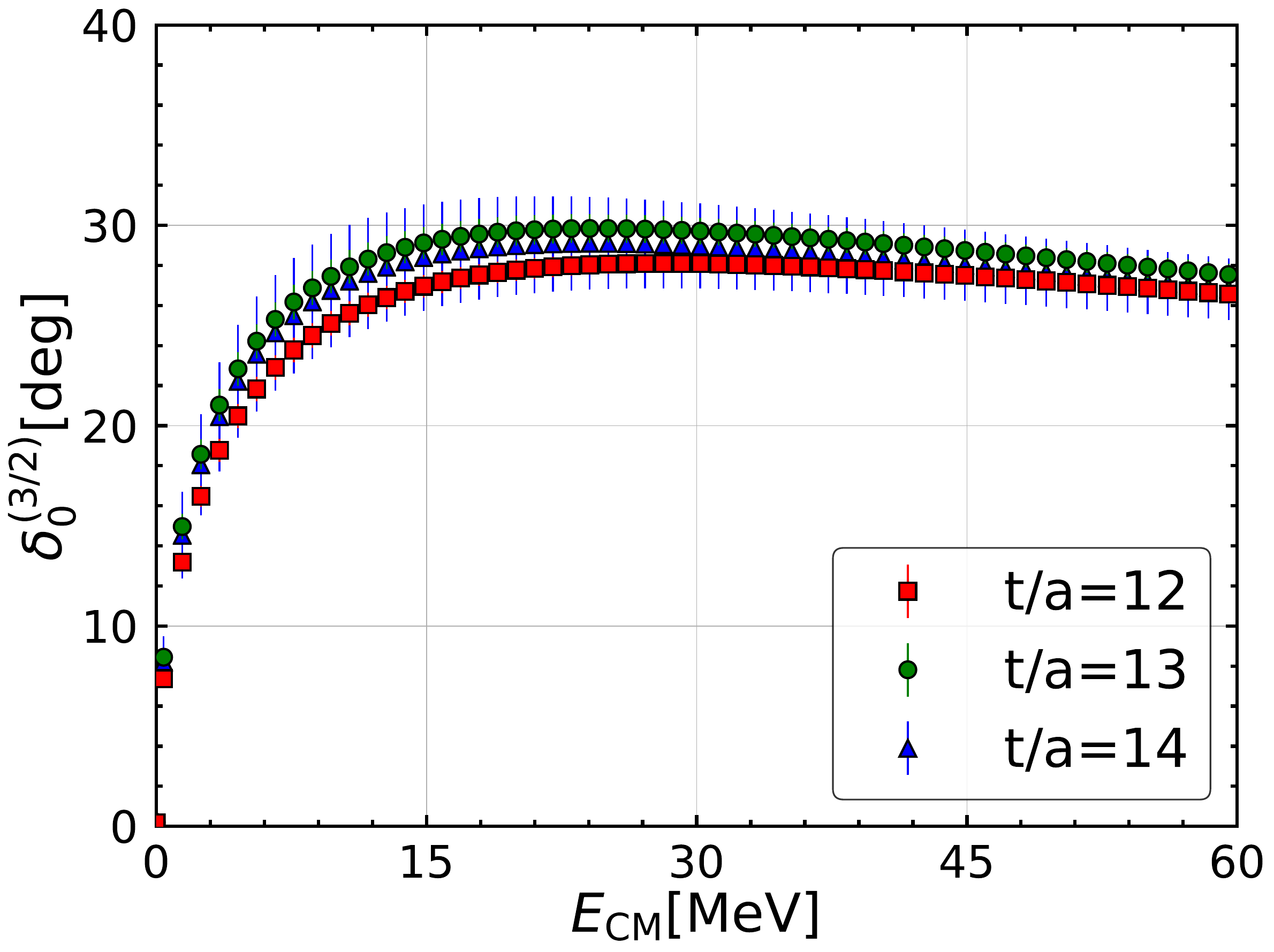}
 %\vskip -2.5cm
 \caption{(Left) The $N-\phi$ potential in the ${}^4S_{3/2}$ channel at $t/a=12$ (red squares), 13 (green circles) and 14 (blue triangle).
 (Right) The phase shifts $\delta^{3/2}$ of the $N-\phi$ scattering in the ${}^4S_{3/2}$ channel, obtained by using $V_{\rm fit}(r)$ at three $t/a$.
  Figures taken from Ref.~\cite{Lyu:2022imf}. }
% \vskip -0.5cm 
 \label{fig:pot_Nphi}
\end{figure*}
The potential between nucleon and $\phi$ meson in the ${}^4S_{3/2}$ channel  was calculated on configurations at 
$m_\pi\simeq 146$ MeV\cite{Lyu:2022imf}, %almost physical pion\cite{Lyu:2022imf}.
where  there are many channels below the $N\phi({}^4S_{3/2})$ threshold at 2092 MeV.
Hoever, contributionrs from 
$\Lambda K ({}^2D_{3/2}) $ at 1665 MeV and  $\Sigma K ({}^2D_{3/2}) $ at 1747 MeV
are kinematically suppressed at low energy due to their $D$-wave nature, and those from 
$\Lambda\pi K$ (1811 MeV), $\Sigma\pi K$ (1893 MeV) and $\Lambda\pi\pi K$ (1957 MeV) are suppressed due to their small phase spaces.
Therefore the single channel analysis of the $N$--$\phi$ potential was employed as the first trial. 

Fig.~\ref{fig:pot_Nphi} (Left) shows the $N-\phi$ potential in the ${}^4S_{3/2}$ channel, which is attractive all distance with a long range tail.
As in the case of the $D^* D$ potential, the long range tail is consistent with (Yukawa)$^2$, suggesting that the 2-pion exchange appears rather universally.
As before, the total potential is fitted by two Gaussians plus (Yukawa)$^2$ as
\begin{eqnarray}
V_{\rm fit}(r;m_\pi) &=&  \sum_{i=1,2} a_i e^{-(r/b_i)^2} + \tilde a_3 m_\pi^4 \left( 1-  e^{-(r/b_3)^2} \right)^2 \left( {e^{-m_\pi r} \over r}\right)^2,
\label{eq:fit_Nphi}
\end{eqnarray}
where an extra $m_\pi$ dependence is introduced by denoting $a_3=\tilde a_3 m_\pi^4$.
Fig.~\ref{fig:pot_Nphi} (Right) represents the $N-\phi$ scattering phase shift in the ${}^4S_{3/2}$ channel, which leads to the scattering length 
$a_0^{3/2} =1.43(23)({}^{+36}_{-06})$ fm. This value is slightly reduced as $a_0^{3/2} \simeq 1.25$ fm if we use $m_\pi=138.0$ MeV (isospin averaged physical pion mass)  in $V_{\rm fit}(r;m_\pi)$ instead of $m_\pi=146.4$ MeV in the simulation.

\begin{figure*}[htb]
\centering
%\vskip -2.5cm
  \includegraphics[height=5.8cm]{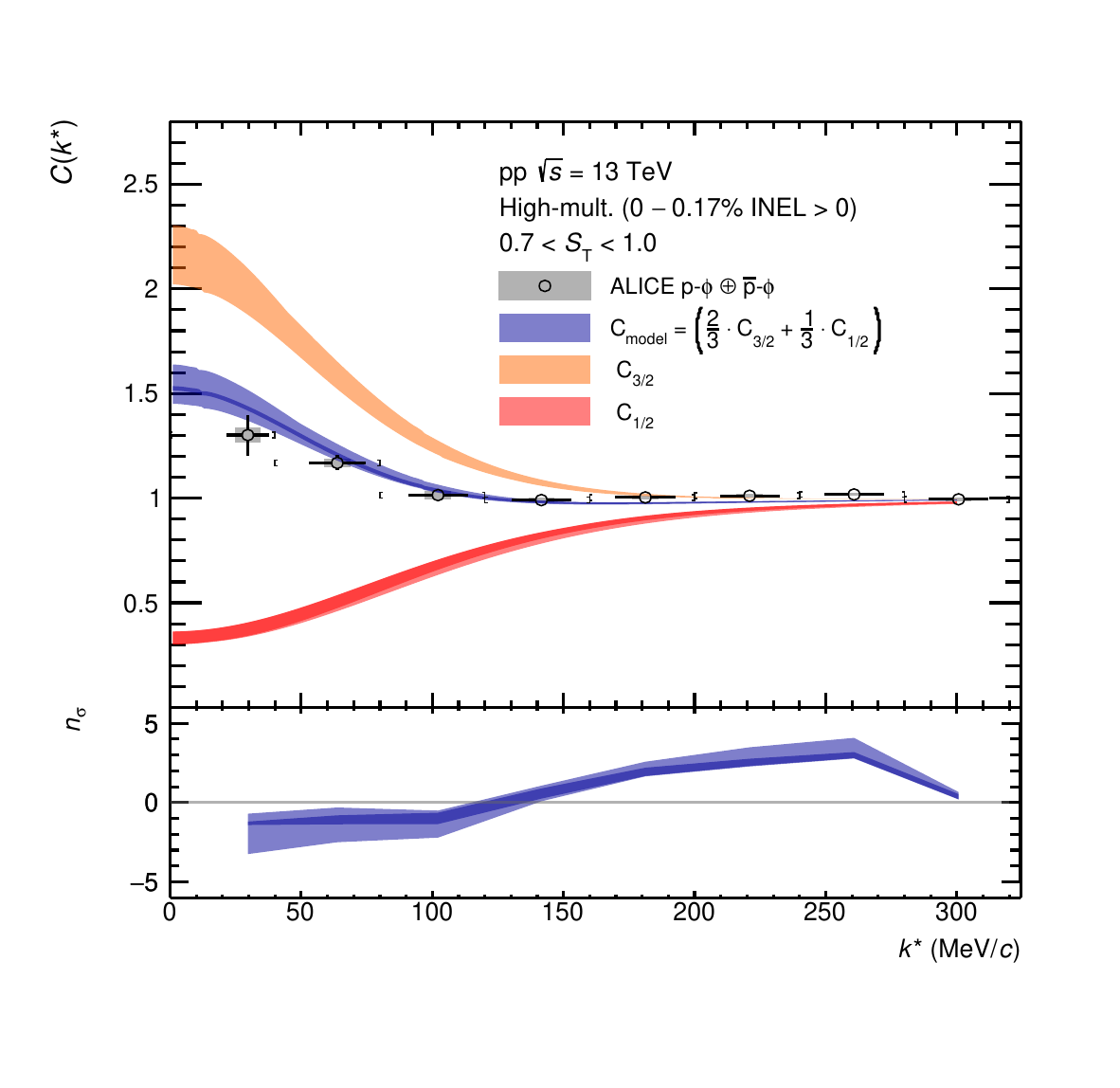}
   \includegraphics[height=5.4cm]{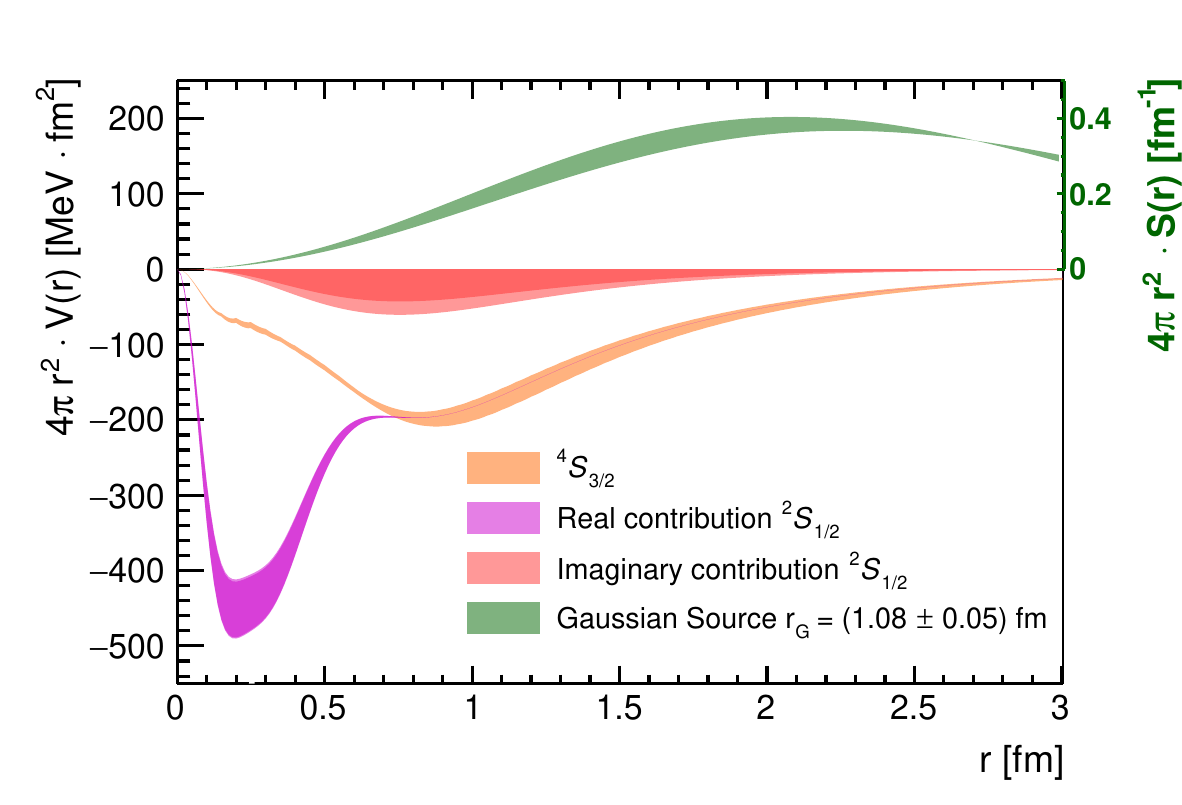}
 %\vskip -2.5cm
 \caption{(Left) The $p-\phi$ correlation functions, measured by the ALICE collaboration\cite{ALICE:2021cpv} (gray shaded squares),
 the ${}^4S_{3/2}$ contribution $C_{3/2}$ calculated with the lattice potential $V_{\rm fit}$,
 the ${}^2S_{1/2}$ contribution $C_{1/2}$ extracted by the fit (red band) and
the  total contribution $2 C_{3/2}/3 + C_{1/2}/3$ (blue band). 
The lower panel is the number of standard deviations between ALICE data and the total contribution.
 (Right) $4\pi r^2 V_{N\phi}^{3/2}(r)$ in the ${}^4S_{3/2}$ channel as a function of $r$ (orange band), together with 
 the real part (violet band) and the imaginary part (red band) of  $4\pi r^2 V^{(1/2)}(r)$ in the ${}^2S_{1/2}$ channel extracted by the fit.
  Figures taken from Ref.~\cite{Chizzali:2022pjd}. }
% \vskip -0.5cm 
 \label{fig:corr_Nphi}
\end{figure*}

Combining the lattice $N\phi ({}^4S_{3/2})$ potential with the $p-\phi$ correlations measured by the ALICE collaboration,
the  $N\phi ({}^2S_{1/2})$ potential was determined\cite{Chizzali:2022pjd}.
Fig.~\ref{fig:corr_Nphi} (Left) shows  the $p-\phi$ correlations measured by the ALICE collaboration\cite{ALICE:2021cpv} (gray shaded squares),
while the contribution $C_{3/2}$ from the ${}^4S_{3/2}$ channel  is estimated by using the lattice potential $V_{\rm fit}$ (orange band).
Since the difference between the experimental data and the  ${}^4S_{3/2}$ contribution represents the contribution $C_{1/2}$ from  the ${}^2S_{1/2}$ channel,
the $N\phi ({}^2S_{1/2})$ potential was determined by assuming the form that
 \begin{eqnarray}
V_{N\phi}^{(1/2)}(r) &=& \beta \sum_{i=1,2} a_i e^{-(r/b_i)^2} +  \left( 1-  e^{-(r/b_3)^2} \right)^2 \left[ \tilde a_3 m_\pi^4 \left( {e^{-m_\pi r} \over r}\right)^2
+i\gamma   {e^{-2m_K r}\over m_K r^2}\right], %\\
\label{fig:VPhi1/2}
\end{eqnarray}
where $\beta,\gamma$ are fit parameters while $a_i, b_i$ are fixed to values for the  $N\phi ({}^4S_{3/2})$ potential, and 
the last term with a pure imaginary coefficient represents the 2nd order kaon exchange with the kaon mass $m_K$.
The ${}^2S_{1/2}$ contribution $C_{1/2}$  (red band) and the total contribution, the weighted average of both $2 C_{3/2}/3 + C_{1/2}/3$  (blue band), 
are also plotted in Fig.~\ref{fig:corr_Nphi} (Left). 
The fit  \eqref{fig:VPhi1/2} works reasonably well.

 Fig.~\ref{fig:corr_Nphi} (Right) represents a comparison among potentials in the form of $4\pi r^2 V(r)$ as a function of $r$,
 where $V=V_{N\phi}^{(3/2)}$ (orange band), $V= {\rm Re} V_{N\phi}^{(1/2)}$ (violet band) and $V= {\rm Im} V_{N\phi}^{(1/2)}$ (red band),
 while the source function $4\pi r^2 S(r)$ is given by the green band.
This comparison  suggests that the attraction at short distance in the ${}^2S_{1/2}$ channel, obtained from the fit to the experimental data,  is stronger than the one in the  ${}^4S_{3/2}$ channel.
The stronger attraction in the $N\phi({}^2S_{1/2})$ leads to
\begin{equation}
{\rm Re}\ a_0^{1/2} = -1.47 ({}^{+0.44}_{-0.37}) ({}^{+0.14}_{-0.17}) \ {\rm fm}, \quad
{\rm Im}\ a_0^{1/2} = 0.00 ({}^{+0.26}_{-0.00}) ({}^{+0.15}_{-0.00}) \ {\rm fm}
\end{equation}
for the scattering length, which indicates an existence of one $N\phi$ bound state, whose estimated binding energy reads
$E_B = 14.7 \sim 56.6$ MeV.
For a further conformation in future, it is important to extract $V_{N\phi}^{(1/2)}$ directly by lattice calculations, where the coupled channel analysis among $N\phi$, $\Sigma K$ and $\Lambda K$ in ${}^2S_{1/2}$ is required.

\section{Summary}
There has been a serious ``NN controversy'' in lattice QCD, that previous studies from the FV method claimed NN systems at heavy pion masses are bound while the HAL QCD method claimed they are unbound.
The cross-checks between the  FV method  and the HAL QCD  methods, as well as those within the FV  method (and within the HAL QCD method), have been performed and cleared the situation:
It is now a consensus among the lattice community that
no bound deuteron and dineutron appear at heavy pion masses (at least at the given lattice spacing).
The previous claims of bound states of three or more baryons in the FV method\cite{Yamazaki:2012hi,NPLQCD:2012mex}
are also very likely to be false since the results relied on the same setups and method
which lead to incorrect conclusions for two baryon bound states.
Accordingly, studies based on the existence of these bound states in the old FV method
(e.g., EFT studies, matrix element studies of nuclei) also lost their foundation. 
 
 The H-dibaryon seems a ``virtual'' state near the $N\Xi$ threshold at  $m_\pi\simeq 146$ MeV and
 lattice $N\Xi$ interactions are compared with LHC data obtained by heavy ion collisions. 
 The $N\Omega$ dibaryon is predicted  and $p\Omega^-$ interactions are calso compared with the LHC data.
 The tetra-quark  state $T_{cc}$  and the $N-\phi$ interaction are investigated also at almost physical pion mass.
 Both show the 2-pion exchange tail at long distances and results derived from these interactions are compared with  experiments.
 More results at not only $m_\pi\simeq 146$ MeV but also $m_\pi\simeq 137$ MeV\cite{Aoyama:2024cko} will be expected.

\end{document}